\documentclass[aps]{revtex4}
\usepackage{amsbsy}
\usepackage{graphicx,color}
\usepackage{amsfonts}
\usepackage{amsmath}
\usepackage{amssymb}

\newcommand{\vect}{\mathbf}

\newcommand{\ds}{ _{\downarrow}}
\newcommand{\us}{ _{\uparrow}}
\newcommand{\up}{\uparrow}
\newcommand{\down}{\downarrow}
\begin{document}
\title{Particle distribution tail and related energy formula}
\author{R. Combescot$^{(a),(b)}$, F. Alzetto$^{(a)}$ and  X. Leyronas$^{(a)}$}
\address{(a) Laboratoire de Physique Statistique,
 Ecole Normale Sup\'erieure,
24 rue Lhomond, 75231 Paris Cedex 05, France}
\address{(b) Institut Universitaire de France}
\date{Received \today}
\pacs{PACS numbers : 03.75.Kk,  05.30.-d, 47.37.+q, 67.90.+z }

\begin{abstract}
We present a simple derivation of the energy formula found by Tan, relative to the single channel hamiltonian
relevant for ultracold Fermi gases. This derivation is generalized to particles with different masses, to arbitrary mixtures, and
to two-dimensional space. We show how, in a field theoretical approach, the $1/k^4$ tail in the momentum distribution
and the energy formula arise in a natural way. As a specific example, 
we consider quantitative calculations of the energy, from different formulas within the ladder diagrams approximation in the normal state.
The comparison of the results provides an indication on the quality of the approximation.
\end{abstract}
\maketitle
\section{INTRODUCTION}

The remarkable progress in the field of ultracold atomic gases has provided access to a number of systems
which may display quite new physical properties. One of the most  striking example is provided by ultracold
fermionic gases \cite{gps} and the BEC-BCS crossover. While the physics of Bose-Einstein condensates (BEC)
is known from superfluid $^4$He (and from ultracold bosonic atomic gases), and the one of BCS  condensates
from superconductors, fermionic gases provide systems which display a smooth continuous evolution between
these two extremes. This is made possible by the existence of Feshbach resonances which allow to control the
value of the scattering length $a$ merely by changing the applied magnetic field. For ultracold gases the kinetic
energy of the atoms is so small that only s-wave scattering is relevant, and it is fully characterized by the
scattering length. Since Pauli principle forbids s-wave scattering between identical atoms, the scattering length corresponds
in most of the experiments performed so far to scattering between atoms in different hyperfine states of a same
element. These hyperfine states are often called for convenience "spin up" and "spin down". 
In the case of a wide Feshbach resonance (as it occurs for example for $^6$Li and $^{40}$K), the closed channel responsible
for the Feshbach resonance may be omitted and the system is described by a single open channel hamiltonian,
where the interaction is characterized by the single parameter $a$. With the densities produced in experiments
this interaction has quite a small range compared to the interatomic distance.
This single channel hamiltonian is particularly interesting because it is at the same time very simple and highly non trivial, and moreover it is realized with an excellent
precision in these ultracold fermionic gases. Hence we may hope, in getting full control and understanding of this hamiltonian, to gain knowledge which may be
applicable to more complex hamiltonians, such as those encountered in condensed matter physics or in quark matter \cite{casal}.

In this context a simple general expression for the energy of a system described by this hamiltonian has been found by Tan \cite{tan1}, 
which involves only the momentum distribution $n_{\sigma}(k)$ of the particles
together with their large $k$ behaviour. However the details raise unanswered mathematical difficulties.
Nevertheless the expression can be checked in limiting cases (see Appendix \ref{A} for details).
This expression is of high interest since it is quite non trivial. It is directly related to the well-known problem
that, for a contact interaction, the kinetic energy presents a formal divergence, because the
momentum distribution $n_{\sigma}(k)$ behaves as $1/k^4$ for large momentum
(as found for example in the perturbative calculation of Belyakov \cite{bely}),
a feature merely linked to two-body physics. Naturally there
is no divergence in the energy itself, because the interaction energy comes in to
compensate this feature from the kinetic energy. This is easily seen from a simple finite range interaction
model (for example the square well potential), with range $r_0$. 
The momentum distribution decreases more rapidly than $1/k^4$, for momentum
beyond $1/r_0$ which acts as a cut-off in momentum space. The apparently
singular situation is found when one lets $r_0 \rightarrow 0$, which should be taken
as the definition of the contact interaction. In the formula found by Tan, the total energy appears essentially as the kinetic energy
with this divergence problem removed. Indeed the high momentum part, responsible for
the divergence, is subtracted out and an additional
explicit contribution which contains the scattering length appears.

This matter has been taken up recently by Braaten and Platter \cite{brpl} who made use of the operator product expansion developped by K. G. Wilson.
They have in this way derived the energy relation in a compact and formal way. 
In contrast with Tan approach the interaction comes in explicitly.
They have also obtained the expression of the known \cite{LLMQ} adiabatic relation for this contact potential hamiltonian.
This convenient expression in terms of the scattering length and of 
the coefficient of the $1/k^4$ tail in the momentum distribution has been pointed out by Tan \cite{tan2}.
Quite recently it has been used \cite{wtc} to obtain the number of closed-channel molecules in the two-channel model.
Also very recently it has been derived and studied in details by Zhang and Leggett \cite{zhlegg}.

In this paper we present first a derivation of this energy formula, which is simple, explicit, fairly short and avoids unnecessary and uncontrolled
mathematical complications. It is in line with Tan approach, in that it deals only with the kinetic energy. This is done in a careful way
in order to avoid divergences. Our treatment is similar in spirit to Ref.\cite{zhlegg}. This simple proof allows straightforward 
generalizations to more complicated situations, with
unequal masses, several kinds of particles and two-dimensional space. Next we show how, in a field theoretical approach,
the $1/k^4$ tail in the momentum distribution and the energy formula arise in a natural way. Finally, as a specific example, 
we consider also quantitative calculations within the ladder diagrams approximation in the normal state. There are different ways
to obtain the energy and we can compare the various approximate results, which gives an indication on the quality of the approximation.

\section{Detailed derivation}\label{detder}
In this paragraph we consider the case where we have only two kinds of particles which are ultracold fermions. 
The volume of the system is assumed to be unity.
Namely we have $n\us$ particles with mass $m\us$ and $n\ds$ particles with mass $m\ds$.
We consider directly the case where the masses $m\us$ and $m\ds$ are different since it does not make any problem. 
The positions of the $\up$ particles are denoted ${\bf r}_i$,
while those of the $\down$ particles are ${\boldsymbol{\rho}}_j$.
For these ultracold fermions only s-wave scattering has to be taken into account. Hence only interactions between $\up$ and $\down$ particles 
have to be considered. Generalizations are considered in the next section.
The Hamiltonian reads $H=H_c+{\mathcal V}$ with (we take $\hbar=1$):
\begin{eqnarray}\label{}
H_c = -\frac{1}{2m\us} \sum_{i=1}^{n\us}\;\Delta _{\bf r_i}-\frac{1}{2m\ds} \sum_{j=1}^{n\ds}\;\Delta _{{\boldsymbol{\rho}}_j}
\end{eqnarray}
and
\begin{eqnarray}\label{}
{\mathcal V}= \sum_{i,j}\;V({\bf r}_i - {\boldsymbol{\rho}}_j)
\end{eqnarray}
We assume the interaction potential $V({\bf r})$ to be short-range. For clarity and simplicity we assume that it has a definite range $r_0$ and satisfies
$V({\bf r})=0$ for $r>r_0$. However, just as in standard scattering theory \cite{LLMQ}, the results will hold for  physical short-range potentials where the
interaction decreases rapidly enough with interparticle distance.

We consider an eigenfunction $|\Phi \rangle$ of this Hamiltonian, having for example in mind the ground state wavefunction. 
However this is not necessary and we can as well
consider excited states, which leads to the extension of the results to non zero temperature, as pointed out by Tan \cite{tan1}. 
Let $\Phi(\{{\bf r}_i\},\{{\boldsymbol{\rho}}_j\})$ be the corresponding wavefunction, with proper symmetrization and normalization to unity.
We want to calculate the energy corresponding to this state:
\begin{eqnarray}\label{en1}
E = \langle \Phi| H |\Phi \rangle =  \int\;\,d{\bf r}_i\,d{\boldsymbol{\rho}}_j
\;\Phi^*(\{{\bf r}_i\},\{{\boldsymbol{\rho}}_j\})\left(H_c+{\mathcal V}\right)\Phi(\{{\bf r}_i\},\{{\boldsymbol{\rho}}_j\})
\end{eqnarray}
where $d{\bf r}_i\,d{\boldsymbol{\rho}}_j$ is for $\prod_{i,j}\,d{\bf r}_i\,d{\boldsymbol{\rho}}_j$ with $i=1,\cdots,n\us$ and $j=1,\cdots,n\ds$.

We follow the basic principle put forward by Tan \cite{tan1}. In Eq.(\ref{en1}) the potential $V({\bf r})$ comes in only in the regions of the integration domain
where $|{\bf r}_m - {\boldsymbol{\rho}}_n|<r_0$, where $m=1,\cdots,n\us$ and $n=1,\cdots,n\ds$. 
The overlap between these regions is negligible for dilute gases where $r_0$ is small compared with the mean
interparticle distance $d$. There are $n\us\,n\ds$ such regions. Compared to the total integration domain, the domain of these regions is of order 
$(r_0/d)^3$. For equal populations $n\us=n\ds \equiv k_F^3/6\pi ^2$, this would be of order $(k_Fr_0)^3$. Since we have $H\Phi=E\Phi$ also inside
these regions, and that $\Phi$ inside these regions has no singular behaviour (this can be checked explicitly since this behaviour is obtained
from a one-body Schr\"{o}dinger equation, as we use below), we may neglect the contribution of these regions. Actually this has to be done for
consistency since \cite{zhlegg} keeping only s-wave scattering for ultracold gases implies that terms of order $(k_Fr_0)^2$, corresponding to higher
angular momenta, are neglected. Hence the energy is simply obtained by calculating the kinetic energy outside these regions:
\begin{eqnarray}\label{}
E =   \int_{|{\bf r}_m - \boldsymbol{\rho}_n|>r_0}\;\,d{\bf r}_i\,d{\bf \rho}_j\;\Phi^*(\{{\bf r}_i\},\{\boldsymbol{\rho}_j\}) H_c \Phi(\{{\bf r}_i\},\{\boldsymbol{\rho}_j\})
\end{eqnarray}

We introduce now the Fourier transform $\varphi(\{{\bf k}_i\},\{{\bf q}_j\})$ of $\Phi(\{{\bf r}_i\},\{\boldsymbol{\rho}_j\})$ with respect to all variables:
\begin{eqnarray}\label{}
\Phi(\{{\bf r}_i\},\{\boldsymbol{\rho}_j\})=  \sum_{\{{\bf k}_i\},\{{\bf q}_j\}} \;e^{i \sum_{i}{\bf k}_i.{\bf r}_i + i \sum_{j}{\bf q}_j.{\boldsymbol{\rho}}_j}\;
\varphi(\{{\bf k}_i\},\{{\bf q}_j\})
\end{eqnarray}
where we use the notation $ \sum_{\bf k} \equiv  (2\pi )^{-3}\int d{\bf k}$. The one-particle density distributions $n\us (k)$ and $n\ds (k)$ are given by:
\begin{eqnarray}\label{eqnup}
n\us (k_1)&=&n\us\,  \sum_{\{{\bf k}_i,i\neq 1\},\{{\bf q}_j\}}\,|\varphi(\{{\bf k}_i\},\{{\bf q}_j\})|^2  \\ \nonumber
n\ds (q_1)&=&n\ds\,  \sum_{\{{\bf k}_i\},\{{\bf q}_j,j\neq 1\}}\,|\varphi(\{{\bf k}_i\},\{{\bf q}_j\})|^2
\end{eqnarray} 
This leads to:
\begin{eqnarray}\label{eqetf}
E &=& \sum_{\{{\bf k}'_i\},\{{\bf q}'_j\}} \;\varphi^*(\{{\bf k}'_i\},\{{\bf q}'_j\})\, \\ \nonumber
&&\sum_{\{{\bf k}_i\},\{{\bf q}_j\}}\left[ \frac{1}{2m\us}\sum_{M=1}^{n\us} k_M^2
 + \frac{1}{2m\ds}\sum_{N=1}^{n\ds} q_N^2 \right]\,\varphi(\{{\bf k}_i\},\{{\bf q}_j\})
\int_{|{\bf r}_m - \boldsymbol{\rho}_n|>r_0}\;d{\bf r}_i\,d\boldsymbol{\rho}_j\;e^{i \sum_{i}({\bf k}_i-{\bf k}'_i).{\bf r}_i + 
i \sum_{j}({\bf q}_j-{\bf q}'_j).\boldsymbol{\rho}_j}
\end{eqnarray}
If we did not have the restrictions $|{\bf r}_m - \boldsymbol{\rho}_n|>r_0$ in the last integral, it would give a factor proportional to $\prod_i \delta({\bf k}_i-{\bf k'}_i)
\prod_j \delta({\bf q}_j-{\bf q'}_j)$ and we would find the standard
expression for the kinetic energy:
\begin{eqnarray}\label{eqkinen}
E_c=\frac{1}{2m\us}  \sum_{{\bf k}_1} \,k_1^2\,n\us (k_1)+\frac{1}{2m\ds}  \sum_{{\bf q}_1}\,q_1^2\,n\ds (q_1)
\end{eqnarray}
which displays a divergence for large wave vector. 
Hence the restriction is crucial in order to avoid the divergence. 

However, since we want
to have the standard expression for the kinetic energy appearing, we will write:
\begin{eqnarray}\label{eqsep}
\int_{|{\bf r}_m - \boldsymbol{\rho}_n|>r_0}d{\bf r}_i\,d\boldsymbol{\rho}_j=
\int d{\bf r}_i\,d\boldsymbol{\rho}_j- \sum_{m,n}\int_{|{\bf r}_m - \boldsymbol{\rho}_n|<r_0}d{\bf r}_i\,d\boldsymbol{\rho}_j
\end{eqnarray}
where for simplicity we have not written explicitely the exponential integrand appearing in Eq.(\ref{eqetf}).
As long as all the $k_i$ and $q_j$ integrations are not performed in Eq.(\ref{eqetf}), no singularity 
appears so we can handle the various terms in Eq.(\ref{eqsep})
separately. As we have just mentionned the first one leads to the standard kinetic energy expression, so we have just to handle the other terms
quite carefully since the limit $r_0 \rightarrow 0$ is singular. Note that, in the regions $|{\bf r}_m - \boldsymbol{\rho}_n|<r_0$, 
we will deal with the analytic continuation of the
expression of the wavefunction for $|{\bf r}_m - \boldsymbol{\rho}_n|>r_0$ which we will discuss below. 
This is naturally quite different from its physical value in these regions,
since, for these last one, the interaction $V({\bf r})$ has to be taken into account.

Let us consider the term $m=n=1$ in Eq.(\ref{eqsep}). In Eq.(\ref{eqetf}) only the terms $(1/2m\us) k_1^2
 + (1/2m\ds) q_1^2$ in the bracket will be relevant, and for now on we consider only these ones.
Basically we will deal carefully with this two-body  problem.
Accordingly it is convenient to think, as long as we are not back to the many-body problem, that only these two particles scatter.
Equivalently we could handle all the $\{m,n\}$ terms simultaneously, but this is much more cumbersome presentation.
Since they are more convenient to express the condition $|{\bf r}_1 - \boldsymbol{\rho}_1|<r_0$, we introduce
the relative position ${\bf r}={\bf r_1}-\boldsymbol{\rho}_1$ and the center of mass position ${\bf R}=(m\us{\bf r_1}+m\ds\boldsymbol{\rho}_1)/M$,
together with their conjugate variables the total momentum ${\bf K}={\bf k}_1+{\bf q}_1$ and the
relative momentum ${\bf k}=(m\ds {\bf k}_1-m\us {\bf q}_1)/M$ with $M=m\us+m\ds$.
We have:
\begin{eqnarray}\label{eqch}
\frac{1}{2m\us}k_1^2
 + \frac{1}{2m\ds} q_1^2=\frac{1}{2\mu }k^2 + \frac{1}{2M} K^2
\end{eqnarray}
where $\mu =m\us m\ds/(m\us+m\ds)$ is the reduced mass. The difficulties arise from the $k^2/(2\mu)$ term in the large $k$ limit.

It is convenient for the discussion to introduce the Fourier transform ${\bar \varphi}({\bf k},{\bf K},\{{\bf k}_2...\},\{{\bf q}_2...\})$
of the wavefunction with respect to the variables ${\bf r}$ and ${\bf R}$, instead of ${\bf r}_1$ and ${\boldsymbol{\rho}}_1$. It is merely given by:
\begin{eqnarray}\label{eqphi}
{\bar \varphi}({\bf k},{\bf K},\{{\bf k}_2...\},\{{\bf q}_2...\})=\varphi({\bf k}+\frac{m\us}{M} {\bf K},-{\bf k}+\frac{m\ds}{M} {\bf K},\{{\bf k}_2...\},\{{\bf q}_2...\})
\end{eqnarray}
Making use, in Eq.(\ref{eqetf}) and Eq.(\ref{eqsep}), of $({\bf k}_1-{\bf k}'_1).{\bf r}_1 + ({\bf q}_1-{\bf q}'_1).{\boldsymbol{\rho}}_1=
({\bf k}-{\bf k}').{\bf r} + ({\bf K}-{\bf K}').{\bf R}$ , together with $d{\bf r}_1\,d{\boldsymbol{\rho}}_1=d{\bf r}\,d{\bf R}$ and $d{\bf k}_1\,d{\bf q}_1=d{\bf k}\,d{\bf K}$,
we find that, for the $k^2/2\mu $ term in Eq.(\ref{eqch}), we have to deal with:
\begin{eqnarray}\label{eqkterm}
\sum_{{\bf K},{\bf k}{\bf q}_2}  \sum_{{\bf k},{\bf k}'}{\bar \varphi}^*({\bf k}',{\bf K},{\bf k}{\bf q}_2)\,k^2\,
{\bar \varphi}({\bf k},{\bf K},{\bf k}{\bf q}_2)\int_{r<r_0}d{\bf r}\,e^{i({\bf k}-{\bf k'}).{\bf r}}
\end{eqnarray}
where ${\bf k}{\bf q}_2$ is a short-hand for $\{{\bf k}_2,\cdots\},\{{\bf q}_2,\cdots\}$. We have used the fact that integration over all unrestricted position variables gives corresponding $\delta$ functions for the
corresponding wavevector variables, as we have mentionned above.
The problem for large $k$ in this expression is directly linked to the behaviour of the wavefunction for small relative distance $r$, which we consider now.

When the relative distance $r$ is small compared to the mean interparticle distance $d$,
the dependence of the wavefunction $\Phi(\{{\bf r}_i\},\{{\boldsymbol{\rho}}_j\})$ on ${\bf r}$ is given by the solution of the relative motion
of the two-body problem. For ultracold gases the energy corresponding to this motion is nearly zero and the corresponding wavefunction is, for $r>r_0$,
proportional to $\psi(r)$ with $\psi (r) \equiv 1/r-1/a$ where $a$ is the scattering length \cite{gps}. 
This form is actually valid provided $r$ is small compared to a typical particle wavelength,
that is $r \ll d$. Since we have $r_0 \ll d$, there is a range of validity for this form for $r>r_0$. 
We are also interested in the case of large positive $a$, 
where a two-body bound state exists with wavefunction
proportional to $\exp(-r/a)/r$. In this case the above form requires $r \ll a$. There is again a range of validity for this condition for $r>r_0$ since we are in
practice interested in physical situations where the scattering length $a$ is large compared to the potential range $r_0$. Hence for small $r$ we have:
\begin{eqnarray}\label{eqptr}
\Phi(\{{\bf r}_i\},\{{\boldsymbol{\rho}}_j\})=\psi(r){\bar \Phi}({\bf R},\{{\bf r}_2...\},\{{\boldsymbol{\rho}}_2...\})
\end{eqnarray}

 Since in the evaluation of the  $\int_{r<r_0}$ term of Eq.(\ref{eqsep}), we will let $r_0 \rightarrow 0$, only large values of ${\bf k}$
are relevant. Indeed if we considered only bounded
values $k<k_c$, this term would go to zero for $r_0 \rightarrow 0$ (and the $k$ integral would converge because of the cut-off $k_c$).
Hence we have to consider $k \rightarrow \infty$. In this case
the dependence of ${\bar \varphi}({\bf k},{\bf K},\{{\bf k}_2...\},\{{\bf q}_2...\})$ on $k$
is entirely linked to the short distance behaviour on $r$ of $\Phi(\{{\bf r}_i\},\{{\boldsymbol{\rho}}_j\})$ given by Eq.(\ref{eqptr}).
We have in this limit:
\begin{eqnarray}\label{}
{\bar \varphi}({\bf k},{\bf K},\{{\bf k}_2...\},\{{\bf q}_2...\})=\psi_F(k){\bar \Phi}_F({\bf K},\{{\bf k}_2...\},\{{\bf q}_2...\})
\end{eqnarray}
where $\psi_F(k)$ and ${\bar \Phi}_F({\bf K},\{{\bf k}_2...\},\{{\bf q}_2...\})$ are the Fourier transform of $\psi (r)$
and ${\bar \Phi}({\bf R},\{{\bf r}_2...\},\{{\bf \rho}_2...\})$ respectively. 

As it is well known, the Fourier transform of $1/r$ is $4\pi /k^2$. On the other hand, for the calculation of
$k^2\,{\bar \varphi}({\bf k},{\bf K},\{{\bf k}_2...\},\{{\bf q}_2...\})$ in expression (\ref{eqkterm}), we do not have to
take into account the constant $-1/a$ in the wavefunction since it gives zero when we apply on it the kinetic energy operator $\Delta _{\bf r}$
(corresponding to the factor $k^2$). All this Fourier transform calculation can be done quite
carefully by multiplying the wavefunction $\psi (r)$ by a convergence factor $e^{-\eta r}$, 
then letting $\eta \rightarrow 0_+$. This confirms the above results. Finally, when handling expression (\ref{eqkterm}),
we have merely $k^2\,{\bar \varphi}({\bf k},{\bf K},\{{\bf k}_2...\},\{{\bf q}_2...\})=4\pi {\bar \Phi}_F({\bf K},\{{\bf k}_2...\},\{{\bf q}_2...\})$.

Physically:
\begin{eqnarray}\label{}
p(k) \equiv  \sum_{{\bf K},{\bf kq}_2}\,|{\bar \varphi}({\bf k},{\bf K},{\bf kq}_2)|^2
\end{eqnarray}
is the isotropic probability distribution of wavevector ${\bf k}$ in the relative motion of the two $\up$ and $\down$ particles we are considering. 
We have found that its leading
behaviour for $k \rightarrow \infty$ is $p_4/k^4$ with:
\begin{eqnarray}\label{eqp4}
p_{4}=(4\pi )^2\,\sum_{{\bf K},{\bf kq}_2}
\,|{\bar \Phi}_F({\bf K},{\bf kq}_2)|^2
\end{eqnarray}

On the other hand, with respect to ${\bf k}'$ integration in expression (\ref{eqkterm}),
it is more convenient to go back to ${\bf r}$ space through $ \sum_{{\bf k}'}e^{-i{\bf k}'.{\bf r}}\psi_F^*(k')=\psi(r)$, which gives a
factor $\psi(r){\bar \Phi}_F^*({\bf K},\{{\bf k}_2...\},\{{\bf q}_2...\})$.
Since $\psi (r)$ is isotropic, we can perform explicitely the angular ${\bf r}$ integration  $\int d\Omega_{\bf r} e^{i{\bf k}.{\bf r}}=4\pi \sin (kr) / (kr)$.
Finally, for expression (\ref{eqkterm}), we are left with the calculation of $
\int_{0}^{r_0} dr\;r \sin (kr) \psi (r)
$. This $r$ integration is easily performed, leading to:
\begin{eqnarray}\label{}
k \int_{0}^{r_0} dr\,r \sin (kr) \left(\frac{1}{r}-\frac{1}{a}\right)  = 1-\left((1-\frac{r_0}{a})\cos(kr_0) +\frac{1}{ka}\sin(kr_0)\right) \equiv 1-f(k,r_0)
\end{eqnarray}

We write now, with this result, the partial contribution $E_{11k}$ to the energy Eq.(\ref{eqetf}), coming from the term $m=n=1$
in Eq.(\ref{eqsep}) and where we retain only the $k^2/2\mu $ term in Eq.(\ref{eqch}). We obtain:
\begin{eqnarray}\label{eq11}
E_{11k}=-\frac{1}{2\mu }\int \frac{d{\bf k}}{(2\pi )^3}\; \frac{1}{k^2}\, p_4\,\left(1-f(k,r_0)\right)=
-\frac{1}{4\pi ^2 \mu } \int_{0}^{k_c}dk\,p_4+\frac{p_4}{4\pi ^2 \mu } \int_{0}^{\infty}dk\,f(k,r_0)
\end{eqnarray}
In the first term in the right-hand side, we have put a cut-off $k_c$ since this integral diverges when $k_c \rightarrow \infty$.
This divergence will just compensate in the final result the above mentionned divergence in Eq.(\ref{eqkinen}).
Hence the global result will be convergent as expected. 

If in the last integral of Eq.(\ref{eq11}) we were setting $r_0=0$, we would get $f(k,r_0)=1$ and a divergent result. 
Instead this integral is perfectly convergent when we calculate it explicitly for $r_0 \neq 0$ and then
take properly the $r_0 \rightarrow 0$ limit. Indeed we see that it is not divergent because
$f(k,r_0)$ involves oscillatory functions.
The $\cos(kr_0)$ term merely gives a result proportional to $\delta(r_0)$, where $\delta(x)$ is the Dirac distribution. 
Since $r_0 \neq 0$ the contribution of this $\cos(kr_0)$ term is zero.  On the other hand, owing to:
\begin{eqnarray}\label{}
 \int_{0}^{\infty} dx \;\frac{\sin x}{x}=\frac{\pi }{2}
\end{eqnarray}
the $\sin(kr_0)$ term gives a contribution $\pi/2a$ to the integral, leading to a contribution $p_4/(8\pi \mu a) $ to $E_{11k}$. 
As we have mentionned above, we could improve the presentation of our handling of the $k$ integration, to deal
with perfectly defined integrals, by
introducing a convergence factor $\exp(-\eta k)$ and then let $\eta\rightarrow 0_+$. Physically this would correspond to regularize
the wavefunction in the $r\rightarrow 0$ limit. This would confirm our above results.

We have not yet taken into account the $K^2/2M$ term in Eq.(\ref{eqch}), because there is no singular behaviour associated with it. 
The corresponding contribution $E_{11K}$ of the term $m=n=1$ in Eq.(\ref{eqsep}) goes to zero as $r_0 \rightarrow 0$,
as it is obvious directly and can be checked by following the same procedure as above. 
Finally one sees easily that, for all the other kinetic energy terms $(1/2m\us)\sum_{m=2}^{n\us} k_m^2
 + (1/2m\ds)\sum_{n=2}^{n\ds} q_n^2$ in the bracket of Eq.(\ref{eqetf}), the contribution from
the term $m=n=1$ in Eq.(\ref{eqsep}) also goes to zero when $r_0 \rightarrow 0$.
Hence the total contribution $E_{11}$ of the $m=n=1$ term is merely $E_{11}=E_{11k}$.

We rewrite now the sum $E_{11}$ in terms of the variables corresponding to particle $\up$
and $\down$, instead of the relative and center of mass variables. We have:
\begin{eqnarray}\label{}
E_{11}= -\frac{1}{2\mu }\sum_{{\bf k},{\bf K}}\frac{P_4({\bf K})}{k^2}+\frac{p_4}{8\pi \mu a}
\end{eqnarray}
with again a cut-off $k_c$ understood for the summation over ${\bf k}$, and where we have introduced:
\begin{eqnarray}\label{eqPkK}
P_{4}({\bf K})=(4\pi )^2\,\sum_{{\bf kq}_2}\,|{\bar \Phi}_F({\bf K},{\bf kq}_2)|^2
\end{eqnarray}
related to $p_4$ by 
$p_4= \sum_{\bf K}P_4({\bf K})$.
With $1/\mu =1/m\us+1/m\ds$, we have:
\begin{eqnarray}\label{eqE1}
E_{11}= -\frac{1}{2m\us }\sum_{{\bf k},{\bf K}}\frac{P_4({\bf K})}{k^2}
-\frac{1}{2m\ds }\sum_{{\bf k},{\bf K}}\frac{P_4({\bf K})}{k^2}
+\frac{p_4}{8\pi \mu a}
\end{eqnarray}
We make use of Eq.(\ref{eqch}) to go back to the ${\bf k}_1$ and ${\bf q}_1$ variables. 
In the first term we change the summation variables from $\{{\bf K},{\bf k}\}$ to $\{{\bf K},{\bf k_1}={\bf k}+(m\us/M){\bf K}\}$, 
making use of $d{\bf k}\,d{\bf K}=d{\bf k_1}\,d{\bf K}$.
In particular we have $1/k^2=1/({\bf k_1}-(m\us/M){\bf K})^2$. However we can use the identity:
\begin{eqnarray}\label{}
 \int d{\bf r}\left[\frac{1}{({\bf r}-{\bf A})^2}-\frac{1}{({\bf r}+{\bf A})^2}\right]=0
\end{eqnarray}
where ${\bf A}$ is any fixed vector (this result is obvious by  changing ${\bf r}$ into $-{\bf r}$),
to write, $\sum_{\bf k_1}[1/k^2-1/k_1^2]=0$ at fixed ${\bf K}$.
Proceeding in the same way with the second term of Eq.(\ref{eqE1}), we obtain:
\begin{eqnarray}\label{eqfin11}
E_{11}= -\frac{1}{2m\us }\sum_{{\bf k}_1}\frac{p_4}{k_1^2}
-\frac{1}{2m\ds }\sum_{{\bf q}_1}\frac{p_4}{q_1^2}
+\frac{p_4}{8\pi \mu a}
\end{eqnarray}
with a cut-off $k_c$ understood for the summation over ${\bf k}_1$ and ${\bf q}_1$.

In conclusion we have calculated all the contributions coming from the presence of the term $m=n=1$ in Eq.(\ref{eqsep}).
These are just the three terms, proportional to $p_4$, in Eq.(\ref{eqfin11}). To summarize, for the expression
of the energy, we are back to Eq.(\ref{eqetf}) with the restriction $|{\bf r}_1 - {\bf \rho}_1|>r_0$ removed, and the above three $p_4$
terms added. 

We have just to repeat the same argument 
for all the other restrictions $|{\bf r}_i - {\bf \rho}_j|>r_0$. In this way we obtain $n\us n\ds$
analogous $p_4$ terms, which are naturally all equivalent after a change of variables. 
Taking Eq.(\ref{eqkinen}) into account, this leads to the final expression for the energy:
\begin{eqnarray}\label{eqfin}
E= \frac{1}{2m\us }\sum_{{\bf k}}\left[k^2\,n\us (k)-\frac{n_4}{k^2}\right]
+\frac{1}{2m\ds }\sum_{{\bf q}}\left[q^2\,n\ds (q)-\frac{n_4}{q^2}\right]
+\frac{n_4}{8\pi \mu a}
\end{eqnarray}
where $n_4=n\us n\ds p_4 =\lim _{k \rightarrow \infty}k^4 n\us(k)=\lim _{k \rightarrow \infty}k^4 n\ds(k)$. Indeed, from Eq.(\ref{eqnup}),
$n\us(k)$ and $n\ds(k)$ behave in this way for large $k$ since for example ${\bf k}_1$ scatters with all the ${\bf q}_j$,
and each scattering brings a contribution $p_4/k^4$. That is any of the $n\us$
particles scatters with any of the $n\ds$ particles. Hence the summations in Eq.(\ref{eqfin}) are perfectly convergent.
This formula is the simple generalization of  the formula found by Tan \cite{tan1} to the case 
where the two species of involved particles have different masses.

In this derivation we introduced a cut-off to manipulate separately each contribution. This makes an easier presentation
for the derivation. However this is just a convenience. We could avoid it by handling all the terms simultaneously.
The presentation would be much awkward, but we would only deal with well defined convergent integrals,
without any need for a cut-off.

\section{SIMPLE GENERALIZATIONS}

It is first worthwhile to note that the above derivation did not make use of the statistics of the particles. Hence the result is valid
for bosons as well as for fermions. Naturally it is artificial for bosons to consider only scattering between different species,
although it might just happen that this scattering is the dominant one. Nevertheless for bosons there is no reason to exclude
scattering between particles belonging to the same species. If we consider first the case of a single bosonic species,
with $n$ particles of mass $m$ in the unit volume and with
scattering length $a$, we can follow the same procedure as in the preceding section. We can write the equivalent of Eq.(\ref{eqetf})
for a single species, with a restriction $|{\bf r}_i - {\bf r}_j|>r_0$ working now between any of the $n(n-1)/2$ couples of particles.
Taking care of the restrictions in the same way as in the preceding section, we end up with:
\begin{eqnarray}\label{}
E= \frac{1}{2m}\sum_{{\bf k}}\left[k^2\,n (k)-\frac{n_4}{k^2}\right]
+\frac{n_4}{8\pi m a}
\end{eqnarray}
with $n_4=n(n-1) p_4=\lim _{k \rightarrow \infty}k^4 n(k)$.

We can then generalize this result to any mixture of $N$ boson species with $n_i$ particles of mass $m_i$ in the unit volume
($i=1,\cdots,N$) and interaction between species $i$ and $j$ characterized by scattering lengths $a_{ij}$. We find in the same way:
\begin{eqnarray}\label{}
E=  \sum_{i=1}^{N}\frac{1}{2m_i }\sum_{{\bf k}}\left[k^2\,n_i (k)-\frac{n_{4i}}{k^2}\right]
+ \frac{1}{16\pi}\sum_{i,j}\frac{n_{4ij}}{\mu_{ij} a_{ij}}
\end{eqnarray}
where $\mu _{ij}=m_im_j/(m_i+m_j)$ is the reduced mass for the $ij$ scattering, 
$n_{4i}=\lim _{k \rightarrow \infty}k^4 n_i(k)$ and $n_{4i}= \sum_{j}n_{4ij}$. Naturally the formula
is more complex for these mixtures, since it requires the knowledge of the $N(N+1)/2$ constants $n_{4ij}=n_{4ji}$ associated with the $i-j$
scattering. The same result works for ultracold fermionic mixtures, except that we have to set $n_{4ii}=0$ because of Pauli principle.

\section{The 2D case}

While in 1D situations the kinetic energy converges, and indeed in exact solutions of many-body
problems the energy is precisely calculated
by evaluating the kinetic energy, there is in 2D a divergence analogous to the 3D case, except that the divergence
is logarithmic in the 2D case. We show here how the
procedure followed in 3D can be extended to this 2D case. Actually there is not so much difference since the space
dimensionality does not appear in the principle of the procedure. The changes appear only when one comes to pratical
matters. We keep the same notations for the variables, but naturally we have to deal now with two dimensional variables
and integrations.

First the expression of the wavefunction $\psi(r)$ for the relative motion at small distance $r>r_0$ is modified into
$\psi (r) \equiv \ln(a/r)$, which is solution of the 2D equation $\Delta \psi(r)=0$. The length $a$ for which $\psi(a)=0$ plays
the role of the scattering length and is naturally obtained from the interaction potential. The Fourier transform of $\psi(r)$
is $\psi_F(k)=2\pi /k^2$, omitting the irrelevant Fourier transform of $\ln a$. Similarly the large $k$ behaviour of $p(k)$
is $p_4/k^4$, with $(2\pi )^2$ appearing in Eq.(\ref{eqp4}) instead of $(4\pi )^2$. Then the 2D angular integration gives
$ \int_{0}^{2\pi }e^{i{\bf k}.{\bf r}}=2\pi \,J_0(kr)$ where $J_0(x)$ is the first kind Bessel function. Hence, instead of Eq.(\ref{eq11}),
we obtain:
\begin{eqnarray}\label{eq112D}
E_{11k}=-\frac{p_4}{2\mu }\int \frac{d{\bf k}}{(2\pi )^2}\; \, \int_{0}^{r_0}dr\, r\,\ln \frac{a}{r}\, J_0(kr)=
-\frac{p_4}{4\pi \mu } \int_{0}^{k_c}\,\frac{dk}{k} \int_{0}^{kr_0}dx\, x\,\ln \frac{ka}{x}\, J_0(x)
\end{eqnarray}
where in the last expression we have changed to the variable $x=kr$, and we have naturally to take the cut-off $k_c \rightarrow \infty$
at the end of the calculation, when the standard kinetic energy term is included, as we have done in section \ref{detder}.
The logarithmic divergence arising in this standard term is expected to be compensated
by the contributions of all the terms similar to Eq.(\ref{eq112D}). Indeed, integrating by parts, we have:
\begin{eqnarray}\label{eqterma}
\int_{0}^{k_c}\frac{dk}{k} \int_{0}^{kr_0}dx\, x\,\ln x\, J_0(x)=\left[\ln k \int_{0}^{kr_0}dx\, x\,\ln x\, J_0(x)\right]^{k_c}_{0}-
r^2_0 \int_{0}^{k_c}dk\, k\,\ln k \ln(kr_0)\, J_0(kr_0)
\end{eqnarray}
To evaluate the first term we use \cite{gr} (when the integral is not absolutely convergent, we define
it as above with an exponential convergence factor, with an extremely weak decreasing behaviour):
\begin{eqnarray}\label{eqbessel}
 \int_{0}^{\infty}dx\,x^{\mu }\,J_0(x)=2^{\mu }\;\frac{\Gamma(\frac{1+\mu }{2})}{\Gamma(\frac{1-\mu }{2})}
\end{eqnarray}
(where $\Gamma(x)$ is the standard Euler gamma function) together with $\ln x=\lim_{\epsilon \rightarrow 0}(x^{\epsilon }-1)/\epsilon $. 
Note in particular that $\int_{0}^{\infty}dx\, x\, J_0(x)=0$. This gives $\int_{0}^{\infty}dx\, x\,\ln x\, J_0(x)=-1$,
which is also consistent with the fact that the Fourier transform of $\ln r$ is $-(2\pi )/k^2$. Hence we find indeed that
Eq.(\ref{eqterma}) provides the required compensating term to avoid a divergent result. 
The other terms lead to constants, so we may take immediately in their expression the limit
$k_c \rightarrow \infty$. First, except for the prefactor $-p_4/(4\pi \mu )$, the factor of $\ln a$ in the second term of the integral in Eq.(\ref{eq112D}) is:
\begin{eqnarray}\label{}
\int_{0}^{\infty}\frac{dk}{k} \int_{0}^{kr_0}dx\, x\, J_0(x)=\int_{0}^{\infty}\frac{dy}{y} \int_{0}^{y}dx\, x\, J_0(x)=-\int_{0}^{\infty}dy\, y\,\ln y\, J_0(y)=1
\end{eqnarray}
by the change of variable $y=kr_0$ and by integrating by parts. This means that the divergent contribution of the second
term in Eq.(\ref{eq112D}) is proportional to $\ln (k_ca)$, as could be expected from dimensional analysis.

Finally in Eq.(\ref{eq112D}) we have still a contribution proportional to $\ln k$ :
\begin{eqnarray}\label{}
-\int_{0}^{\infty}\frac{dk}{k} \,\ln k\int_{0}^{kr_0}dx\, x\, J_0(x)=\frac{r_0^2}{2}\int_{0}^{\infty}dk\, k\,\ln^2 k \, J_0(kr_0)
\end{eqnarray}
again by integrating by parts. Gathering this term and the second term in Eq.(\ref{eqterma}), we have:
\begin{eqnarray}\label{}
-r^2_0 \int_{0}^{\infty}dk\, k\,\ln k \ln(kr_0)\, J_0(kr_0)
+\frac{r_0^2}{2}\int_{0}^{\infty}dk\, k\,\ln^2 k \, J_0(kr_0)=
-\frac{1}{2}\int_{0}^{\infty}d(kr_0)\, (kr_0)\,\ln^2 (kr_0) \, J_0(kr_0)
\end{eqnarray}
where, in the last step for the term proportional to $\ln^2 r_0$,
we have used again $\int_{0}^{\infty}dx\, x\, J_0(x)=0$. The remaining integral is calculated again
by making use of Eq.(\ref{eqbessel}) and is found to be
$(1/2)\int_{0}^{\infty}dx\, x\,\ln^2 x \, J_0(x)=C-\ln 2 \simeq -0.116$, 
where $C=0.577216\cdots$ is the Euler constant.
Gathering the above results we obtain:
\begin{eqnarray}\label{eq112Da}
E_{11k}=-\frac{p_4}{4\pi \mu } \ln(k_ca)+p_4\,\frac{\ln 2 -C}{4\pi \mu }
\end{eqnarray}

Finally, just as for the 3D case, we want to go back to the variables corresponding to particle $\up$
and $\down$. It may be seen that this does not lead to additional contributions.
This leads us finally to:
\begin{eqnarray}\label{eqfin2d}
E= \frac{1}{2m\us }\lim _{k_c \rightarrow \infty}\left[\sum^{k<k_c}_{{\bf k}}k^2\,n\us (k)-\frac{n_4}{2\pi} \ln(k_ca)\right]
+\frac{1}{2m\ds }\lim _{q_c \rightarrow \infty}\left[\sum^{q<q_c}_{{\bf q}}q^2\,n\ds (q)-\frac{n_4}{2\pi} \ln(q_ca)\right]
+n_4\frac{\ln 2 -C}{4\pi \mu }
\end{eqnarray}
where we recall that $n_4=n\us n\ds p_4 =\lim _{k \rightarrow \infty}k^4 n\us(k)=\lim _{k \rightarrow \infty}k^4 n\ds(k)$.
This result can be checked explicitly in the molecular case (see Appendix A).

\section{FIELD THEORETIC APPROACH}\label{fta}

Let us see now how the above expression for the energy arises in the field theoretic formalism.
To be specific we will restrict ourselves for simplicity to the case of major interest, namely the one of
two fermionic species with equal populations, so that $n\us(k)=n\ds(k) \equiv n(k)$.
Since all the $\up$ and $\down$ quantities are equal, we do not write explicitely
this index. We consider also the non zero temperature case
since it does not make any problem.
We deal first with the case of a normal system, and then extend the results to the superfluid case.

\subsection{Normal state}\label{normst}

We consider first the large $k$ dependence of $n(k)$.
For free fermions at temperature $T$, the density distribution has an exponential tail proportional
to $e^{-k^2/2mT}$. However, as pointed out in Ref. \cite{brpl}, interactions modify this behaviour
and give rise on general grounds to a $1/k^4$ dependence which dominates the exponential tail. 
This is explicit in the weak coupling
domain where the interaction can be treated perturbatively and the scattering length $a$ is small, 
as it has been done by Belyakov \cite{bely} at zero temperature:
\begin{eqnarray}
n(k)=\left(\frac{2}{3\pi }\,k_F a\right)^{2} \;\frac{k_F^4}{k^4}\label{eqnk}
\end{eqnarray}
where $k_F$ is the Fermi momentum $n\us = n\ds = k_F^3/(6\pi ^2)$.

In the general case the distribution $n(k)$ is obtained from the temperature 
Green's function $G({\bf k},i\omega _n)$, where $\omega _n = (2n+1) \pi T$ (with $n$ an integer and $k_B=1$) 
is the Matsubara frequency, by:
\begin{eqnarray}
n(k) = T  \sum_{n} \; G({\bf k},i\omega _n)\; e^{i \omega _n \tau}
\label{eqn}
\end{eqnarray}
where $\tau \rightarrow 0_+$. We separate out in this equation the free particle contribution by writing Dysons's equation:
\begin{eqnarray}\label{eqdys}
G({\bf k},i\omega _n)=G_0({\bf k},i\omega _n)+G_0({\bf k},i\omega _n)\Sigma({\bf k},i\omega _n) G({\bf k},i\omega _n)
\end{eqnarray}
where $G_0$ is the free particle Green's function
$G_0({\bf k},i\omega _n) = [i \omega _n - \epsilon_{\bf k} + \mu]^{-1}$, with $\epsilon_{\bf k} = k^{2}/2m$ 
the free particle kinetic energy, and $\Sigma({\bf k},i\omega _n)$ is the self-energy. The first term in Eq.(\ref{eqdys}) gives in
Eq.(\ref{eqn}) the free particle contribution, namely the Fermi distribution. For large $k$ its exponential tail mentionned above is completely
dominated by the algebraic decay $1/k^4$ that we will obtain. Hence we are left only with the second term. 
For large $k$, implying a large kinetic energy for the particle,
we expect the effect of interaction to be small in the same spirit as the Born approximation in this regime. Since $\Sigma ({\bf k}, i\omega _{n}) $
describes this effect we expect it to be small. Hence we may to lowest order replace in this second term $G$ by $G_0$. On the other hand
we replace, in a standard way \cite{agd}, the summation over Matsubara frequencies by a frequency integration over a contour
$\mathcal C$ encircling the imaginary axis in the anticlockwise direction. 
This leads to the following expression for the dominant contribution to $n(k)$ at large $k$:
\begin{eqnarray}\label{eqnk}
n(k)=-\frac{1}{2i\pi } \int_{\mathcal C} d\omega \,f(\frac{\omega }{T})\,\frac{\Sigma({\bf k},\omega)}{(\omega -\epsilon _k +\mu )^2}
\end{eqnarray}
where $f(x)=1/(e^{x}+1)$ is the Fermi distribution function. The contour can be deformed into the sum of a contour enclosing the
positive frequency ${\mathrm Re}\,\omega >0$ half-plane and another contour enclosing the negative frequency 
${\mathrm Re}\,\omega <0$ half-plane, both being in the clockwise direction. Closing the contour at infinity in the 
${\mathrm Re}\,\omega >0$ half-plane
is allowed by the presence of the Fermi distribution $f(\omega /T)$, and closing it at infinity in the 
${\mathrm Re}\,\omega <0$ half-plane by the presence of the $e^{\omega  \tau}$ factor, which is then omitted since it does
not play any other role.

Now contributions from the double pole at $\omega =\epsilon _k-\mu $ will contain from the Fermi distribution a 
factor $e^{-k^2/2mT}$ which makes them negligible. Similarly $\Sigma({\bf k},\omega)$ has also singularities
with frequencies which are large and positive, when $k$ is large.  The Fermi distribution $f(\omega /T)$ will again
make their contribution exponentially small. On the other hand, as it is shown in details in Appendix \ref{B}, $\Sigma({\bf k},\omega)$
has a pole at $\omega \approx -\epsilon _k$ for large $k$, that is deep in the ${\mathrm Re}\,\omega <0$ half-plane.
Roughly speaking this pole appears since a $\up$ particle with large ${\bf k}$ and $\omega $ will scatter with a $\down$
particle with essentially opposite parameters ${\bf -k}$ and $-\omega $. This happens because, just as in section \ref{detder},
the large values occur for the relative motion, but not for the center of mass motion. Then the $\down$ particle propagator
has a pole at $\omega \simeq -\epsilon _k$, which produces a pole for $\Sigma({\bf k},\omega)$ at  the same frequency.

For this pole, we have merely $(\omega -\epsilon _k +\mu )^2 \simeq (2\epsilon _k)^2
=k^4/m^2$ and $f(-\epsilon _k/T) \simeq 1$.
If we call $R_{\Sigma}$ the corresponding residue of $\Sigma({\bf k},\omega)$, we obtain in this way:
\begin{eqnarray}\label{eqnkgdk}
n(k) \simeq m^2 R_{\Sigma}\,\frac{1}{k^4}
\end{eqnarray}
Hence we see that the $1/k^4$ dependence of $n(k)$ at large $k$ emerges quite naturally in this approach.
We note that our analysis may seem inconsistent, since we have first argued that $\Sigma({\bf k},\omega)$
should be small for large $k$, but then considered a pole of $\Sigma({\bf k},\omega)$, in the vicinity of which
it is large. The justification is that, as long as the contour is far away from the pole, $\Sigma({\bf k},\omega)$
is indeed small everywhere on the contour. Making use of the pole is then a convenient way to calculate the
contour integral.

We consider now the expression of the energy.
It can be written in terms of the Green's function by a standard formula \cite{agd} which bears from
the start a strong analogy with the expression found by Tan \cite{tan1}:
\begin{eqnarray}\label{eqengen}
E= T\sum_{{\bf k},n} \; (i\omega _n +\mu +\epsilon _k)\,G({\bf k},i\omega _n)\; e^{i \omega _n \tau}
\end{eqnarray}
where again $\tau \rightarrow 0_+$. This equation includes an overall factor 2, coming from summation over spin. The essence of this formula is that,
calculating $i\omega _n \,G({\bf k},i\omega _n)$ by Heisenberg equations of motion, one finds
$E_c+2E_{\rm int}$, that is the kinetic energy $E_c$ plus twice the interaction energy $E_{\rm int}$. Adding the kinetic energy $E_c$, 
which gives the $\epsilon _k$ term in Eq.(\ref{eqengen}), leads to $2E=2(E_c+E_{\rm int})$.
In practice Eq.(\ref{eqengen}) is not so useful since one has to deal carefully with the divergent behaviour occuring
for large $\omega _n$ and large $k$, which is quite painful numerically. 
In this respect the number equation Eq.(\ref{eqn}) is much more convenient.

Now, to single out clearly the kinetic energy contribution, we can rewrite Eq.(\ref{eqengen}) as:
\begin{eqnarray}\label{eqenb}
E= 2\,T \sum_{{\bf k},n} \; \epsilon _k\,G({\bf k},i\omega _n)\; e^{i \omega _n \tau} +
T\sum_{{\bf k},n} \; (i\omega _n +\mu -\epsilon _k)\,G({\bf k},i\omega _n)\; e^{i \omega _n \tau}
\end{eqnarray}
The first term is just the kinetic energy $E_c=2 \sum_{{\bf k}}\epsilon _k\,n(k)$. In the second term, which is just $E_{\rm int}$, 
we can again replace the Matsubara frequency
summation by the same contour integrals as above. Moreover the quantity appearing in this summation is just $G_0^{-1}({\bf k},\omega)G({\bf k},\omega)
=1+\Sigma({\bf k},\omega) G({\bf k},\omega)$. However the term $1$ gives a zero contribution to these closed contour integrals.
Hence we are left with:
\begin{eqnarray}\label{eqenc}
E=E_c-\frac{1}{2i\pi }  \sum_{\bf k}\int_{\mathcal C} d\omega \,f(\frac{\omega }{T})\,\Sigma({\bf k},\omega)G({\bf k},\omega)
\end{eqnarray}
We can now analyze the behaviour of the interaction energy term for large $k$, as we have done above for the particle number.
In this range $\Sigma({\bf k},\omega)$ is small so we may replace $G({\bf k},\omega)$ by $G_0({\bf k},\omega)$. Hence we
have to deal with the same expression as in Eq.(\ref{eqnk}), except that $G_0({\bf k},\omega)$ is not squared.
Following the same arguments we obtain for large $k$:
\begin{eqnarray}\label{eqintgdk}
-\frac{1}{2i\pi } \int_{\mathcal C} d\omega \,f(\frac{\omega }{T})\,\Sigma({\bf k},\omega)G({\bf k},\omega)
\simeq -\frac{m R_{\Sigma}}{k^2}
\end{eqnarray}
Hence we have shown explicitely that the large $k$ behaviour of the interaction energy 
comes in to cancel the divergent behaviour of the kinetic energy. Naturally this is the expected result.
Accordingly we may write:
\begin{eqnarray}\label{eqen2}
E = 2  \sum_{\bf k} \left[\frac{k^2}{2m}\,n(k)-\frac{m R_{\Sigma}}{2\,k^2}\right]+ \sum_{\bf k}
\left[\frac{m R_{\Sigma}}{k^2}-\frac{1}{2i\pi } \int_{\mathcal C} d\omega 
\,f(\frac{\omega }{T})\,\Sigma({\bf k},\omega)G({\bf k},\omega)\right]
\end{eqnarray}
where, from Eq.(\ref{eqnkgdk}) and Eq.(\ref{eqintgdk}), both brackets give convergent integrals for large $k$.
The first term is just the kinetic energy from which, according to Eq.(\ref{eqnkgdk}), the large $k$ behaviour has been subtracted. By comparison
with Eq.(\ref{eqfin}), the second term in Eq.(\ref{eqen2}) is just $mR_{\Sigma}/(4\pi a)$, a result by no means obvious.

\subsection{Numerical calculations in the ladder approximation}
We are now in a situation where we can calculate the energy from three different
formulas, as we explain just below. For the exact theory they would naturally give the same result. However
if we use an approximate scheme, the results are expected to be different. The differences can, in some rough way, be seen as a measure of the errors resulting
from the approximation. Hence it is of interest to make such a comparison.
Specifically we have taken as an example the ladder approximation \cite{ppsc,clk} 
at unitarity and we have performed numerical calculations of the energy as a function of
temperature in this approximation.  

We use three different ways to obtain the
energy. Our first expression for the energy is the general one Eqs.(\ref{eqenc},\ref{eqen2}). 
The second one is the energy
formula found by Tan, that is Eq.(\ref{eqfin}) with equal masses. 
Finally we use the non-interacting gas scaling relation $E=-\left(  3/2\right)\Omega$ between the grand potential $\Omega$ and the energy $E$,
which is also valid at unitarity  \cite{jho} due to the lack of any microscopic energy scale.
We calculate the grand potential itself {\it via} the density $n(\mu,T)$ 
(obtained by integration  of $n(k)$) using the thermodynamic
relation $\Omega(\mu,T)/V=-\int_{-\infty}^{\mu}d\mu'\,n(\mu',T)$ (we have used the fact that $\Omega$ goes to $0$ in the high temperature regime $\mu/T\to-\infty$).
As noted above this last method is much more convenient numerically than the two others, since there are no convergence problems for large $k$ in the ${\bf k}$ summation.

\begin{figure}
\rotatebox{270}{\includegraphics[width=.5\linewidth]{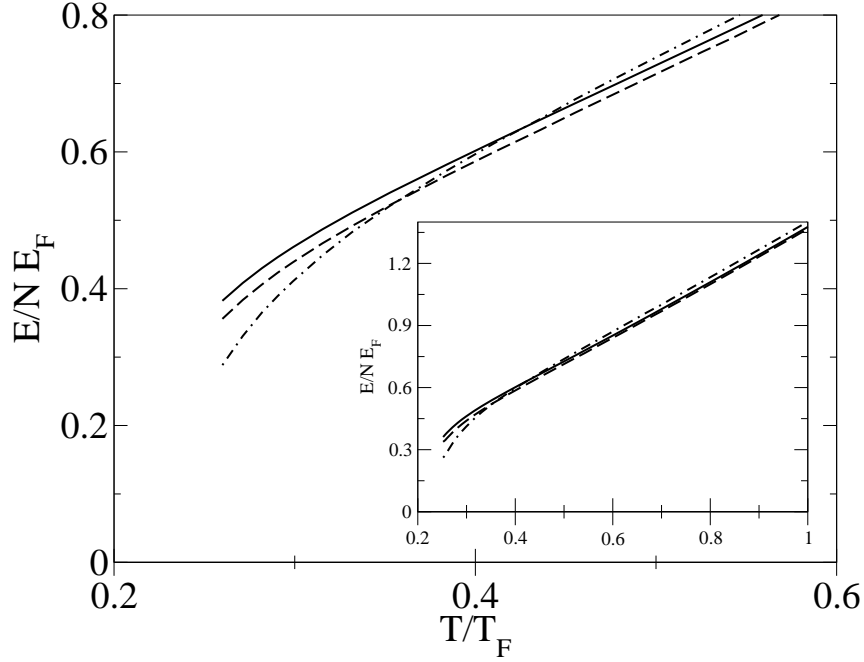}}
\caption{Energy of the gas at unitarity as a function of reduced temperature, obtained from Eq.(\ref{eqen2}) (dashed line),
Eq.(\ref{eqfin}) (dashed-dotted line) and from the scaling relation (full line) $E=-\left(  3/2\right)\Omega$, with the
grand potential $\Omega $ obtained by integrating the particle density. The inset shows the results with an extended scale.}
\label{fig1}
\end{figure}

The results of these numerical calculations are shown in Fig.\ref{fig1}. We have checked that we recover the proper high temperature behavior (virial expansion) with the three methods, as it should be. This common evolution toward the virial result is clearer in the inset, where the results are shown with an extended scale,
but this convergence is fairly slow. The results of the three methods are fairly close, which is coherent with the fact that we expect the ladder approximation to be
reasonably good, mostly toward higher temperatures. Indeed at lower temperature the differences between the three results are growing, mostly
for the result of the Tan formula. This is clearly due to the proximity of the superfluid critical temperature, which occurs at $T/T_F \simeq 0.243$ in this approximation.

The major numerical problem with the use of Eq.(\ref{eqfin}) is to obtain the proper coefficient of
$1/k^4$ in the large momentum behaviour of $n(k)$. This requires in particular to be sure that the
asymptotic regime is reached numerically. The same problem arises in Eq.(\ref{eqen2}) since one
must find numerically that, for large $k$, the kinetic energy is exactly balanced by the interaction energy.
In the present case of the ladder approximation, we can check that the asymptotic regime
as been reached since we have analytical expressions. In order to check numerically this 
large momentum behaviour of the contributions coming into the two brackets of Eq.(\ref{eqen2}), we 
have performed, as we indicate specifically below, an expansion of the occupation number $n(k)$ 
and of the interaction energy term up to  order $k^{-6}$ and $k^{-4}$ respectively. 

As indicated in the above subsection the self-energy is small in the large momentum and frequency limit and we can expand the Green's function in powers of the self-energy. 
Hence we have for the occupation number and the interaction energy term:
\begin{align}
 n \left( k \right) = - \frac{1}{2 \pi i} \int_C d \omega f \left( \frac{\omega}{T} \right) \left( G_0 \left( \vect{k} , \omega \right) + G_0 \left( \vect{k} , \omega \right) \Sigma \left( \vect{k} , \omega \right) G_0 \left( \vect{k} , \omega \right) + \mathcal{O} \left( \Sigma^2 \right) \right)
\label{eq:n_k}\\
 - \frac{1}{2 \pi i} \int_C d \omega f \left( \frac{\omega}{T} \right) \Sigma \left( \vect{k} , \omega \right) G \left( \vect{k} , \omega \right) = - \frac{1}{2 \pi i} \int_C d \omega f \left( \frac{\omega}{T} \right) \left( \Sigma \left( \vect{k} , \omega \right) G_0 \left( \vect{k} , \omega \right) + \mathcal{O} \left( \Sigma^2 \right) \right)
\label{eq:e_int_k}
\end{align}

We start from equation (B1) and following the arguments given above and detailed in Appendix B, 
we neglect the contribution coming from the pole of the Green's function, that gives exponentially small terms, and we only consider the contribution from the cut of the vertex $\Gamma \left( \vect{K}, \Omega \right)$ on the real frequency axis. Following \cite{clk}, we call this self-energy contribution $\Sigma_\Gamma$:

\begin{align}
\Sigma_\Gamma\left(\vect{k},\omega\right)=\frac{1}{\pi}\sum_{\vect{K}}\int_{\Omega_{\text{min}}}^{\infty} d \Omega\,b \left( \frac{\Omega}{T} \right) \text{Im}\,\Gamma \left( \vect{K}, \Omega + i \epsilon \right) G_0 \left( \vect{K} - \vect{k}, \Omega - \omega \right)
\label{eq:sigma_gamma}
\end{align}
with $\Omega_{\text{min}}=\frac{K^2}{4m}-2\mu$, and $\epsilon \rightarrow 0_+$.

The Bose distribution effectively limits the frequency $\Omega$ to be at most a few $T$ and the $K$-integral is therefore effectively bounded. Similarly to what is done in Appendix B, we consider the singularities of the self-energy located at $\omega\approx-\epsilon_{\vect{k}}$.  In the large momentum $\vect{k}$ and large frequency $\omega$ limit, we expand the free Green's function in Eq.(\ref{eq:sigma_gamma}) in power of $ \left(\omega+\epsilon_{\vect{k}}+\mu\right)^{-1}$, leading to:

\begin{align}
\Sigma_\Gamma\left(k,\omega\right)=\sum_{p=0}^{\infty}\frac{C_p\left(k\right)}{\left(\omega+\epsilon_{\vect{k}}+\mu\right)^{p+1}}
\label{eq:sigma_dev}
\end{align}

where 

\begin{align}
 C_p\left(k\right)=-\frac{1}{\pi}\sum_{\vect{K}}\int_{\Omega_{\text{min}}}^\infty d\Omega\,b\left(\frac{\Omega}{T}\right)\text{Im}\,\Gamma\left(\vect{K},\Omega+i\epsilon\right)\times\left(\frac{\vect{k}\cdot\vect{K}}{m}+\Omega-\frac{K^2}{2m}+2\mu\right)^p
\label{eq:c_p}
\end{align}

In this limit, the contribution to the self-energy due to the cut of the vertex $\Gamma \left( \vect{K}, \Omega \right)$ is a sum of poles of order $p+1$ located at $\omega=-\epsilon_{\vect{k}}-\mu$. In Eq.(\ref{eq:n_k}), only the second term is of interest, since the first one gives the Fermi distribution which does not contribute to the algebraic tail. So we use the expansion Eq.(\ref{eq:sigma_dev}) in Eq.(\ref{eq:n_k}) and find the following expansion for large wave vector $k$:

\begin{align}
n(k)\simeq - \frac{1}{2 \pi i} \int_C d \omega f \left( \frac{\omega}{T} \right) \left(G_0\left(\vect{k},\omega\right)\right)^2\Sigma\left(\vect{k},\omega\right)\simeq\sum_{p=0}^\infty \left(p+1\right)\frac{ C_p\left(k\right)}{\left(k^{2}/m\right)^{p+2}}
\end{align}

The first three terms of this expansion give the dominant contribution. Indeed, it is easily checked that the term of order $\Sigma^2$ in Eq.(\ref{eq:n_k}) give contributions at least of order $k^{-8}$.
$C_0$ does not depend on ${\bf k}$ while the ${\bf k}$ dependence of $C_1$ vanishes
after angular integration. For the coefficient $C_2$ we can write
$C_2\left(k\right)=A_2 \frac{k^2}{m} + B_2$.
Finally we find for the particle distribution in the large momentum limit:

\begin{align}\label{nkfin1}
 n\left(k\right)=C_0\frac{m^2}{k^4}+\left(2 C_1 + 3 A_2\right)\frac{m^3}{k^6}+\mathcal{O}\left(\frac{m^4}{k^8}\right)
\end{align}

Comparing with Eq.(\ref{eqnkgdk}) we see that $C_0$ is identical to the coefficient $R_\Sigma$ defined in the above subsection.
 We have checked the numerical convergence in the large $k$ limit by comparing the second term in our expansion Eq.(\ref{nkfin1}) with our numerical results. The coefficients are identical with a precision of less than one percent.

We can perform the same analysis for the interaction energy term in Eq.(\ref{eq:e_int_k}) and we find the following expansion :

\begin{align}
 - \frac{1}{2 \pi i} \int_C d \omega f \left( \frac{\omega}{T} \right) \Sigma \left( \vect{k} , \omega \right) G \left( \vect{k} , \omega \right) = - C_0 \frac{m}{k^2} - \left(C_1 + A_2\right) \frac{m^2}{k^4} + \mathcal{O} \left( \frac{m^3}{k^6} \right)
\end{align}

Here also we have checked the numerical convergence with a precision of less than one percent.

\subsection{Superfluid state}

Compared to the normal state, we have now to modify the Dyson's equation Eq.(\ref {eqdys}) to take into account \cite{agd}
the anomalous self-energy $\Delta ({\bf k},i\omega _n)$
and the anomalous propagator $F^{+}({\bf k},i\omega _n)$. We have:
\begin{eqnarray}\label{eqdysona}
G({\bf k},i\omega _n)&=&G_0({\bf k},i\omega _n)+G_0({\bf k},i\omega _n) 
\Sigma ({\bf k},i\omega _n) G({\bf k},i\omega _n) +G_0({\bf k},i\omega _n) \Delta ({\bf k},i\omega _n) F^{+}({\bf k},i\omega _n) \\
F^{+}({\bf k},i\omega _n)&=&G_0(-{\bf k},-i\omega _n) \Sigma (-{\bf k},-i\omega _n) F^{+}(p) -
G_0(-{\bf k},-i\omega _n) \Delta^{*} ({\bf k},i\omega _n) G({\bf k},i\omega _n) 
\label{eqdysonb}
\end{eqnarray}
Again for large $k$, we have to lowest order $G({\bf k},i\omega _n) \simeq G_0({\bf k},i\omega _n)$ and $\Sigma ({\bf k},i\omega _n)$
small. For such a large $k$ we expect this normal self-energy $\Sigma ({\bf k},i\omega _n)$ to behave as in the normal state, discussed in the preceding
subsection. On the other hand the anomalous self-energy is expected to go to a constant, without any  singular behaviour. 
This implies from Eq.(\ref{eqdysonb}) that $F^{+}({\bf k},i\omega _n)$ is small. In this case the first term in the right-hand side
of Eq.(\ref{eqdysonb}) is negligible and from the second term we have:
\begin{eqnarray}\label{}
F^{+}({\bf k},i\omega _n)& \simeq & -
G_0(-{\bf k},-i\omega _n) \Delta^{*} ({\bf k},i\omega _n) G_0({\bf k},i\omega _n)
\end{eqnarray}
Substitution into Eq.(\ref{eqdysona}) gives:
\begin{eqnarray}\label{}
G({\bf k},i\omega _n)& \simeq &G_0({\bf k},i\omega _n)+G_0^2({\bf k},i\omega _n) \left[
\Sigma ({\bf k},i\omega _n) -G_0(-{\bf k},-i\omega _n) |\Delta ({\bf k},i\omega _n)|^2\right]
\end{eqnarray}
Hence, in addition to the pole at $\omega \approx -\epsilon _k$ coming as discussed above in $\Sigma ({\bf k},i\omega _n)$,
the last term gives also a pole at the same place from the explicit factor $G_0(-{\bf k},-i\omega _n)$. This leads to the large $k$
behaviour:
\begin{eqnarray}\label{}
n(k) \simeq m^2\left[ R_{\Sigma}+|\Delta _{\infty}|^2\right]\,\frac{1}{k^4}
\end{eqnarray}
where $\Delta _{\infty}=\lim _{k \rightarrow \infty}\Delta ({\bf k},-\epsilon _k)$. Actually the additional term is well-known in the case of
BCS theory which gives indeed, with standard notations, $n_k=v_k^2=(1/2)(1-\xi_k/E_k) \simeq m^2\,\Delta ^2/k^4$ for large $k$.

Going now to the expression for the energy, there is naturally no modification in Eq.(\ref{eqenb}). On the other hand we have now,
in the second term of Eq.(\ref{eqenb}) involving $G_0^{-1}G$, to make use of the appropriate Dyson's equation Eq.(\ref{eqdysona}).
This leads, instead of Eq.(\ref{eqenc}), to:
\begin{eqnarray}\label{eqenc1}
E=E_c-\frac{1}{2i\pi }  \sum_{\bf k}\int_{\mathcal C} d\omega \,f(\frac{\omega }{T})\,\left[\Sigma({\bf k},\omega)G({\bf k},\omega)
+\Delta ({\bf k},i\omega) F^{+}({\bf k},i\omega)\right]
\end{eqnarray}
Now the large $k$ analysis goes essentially as in the normal state, and is similar to the one for the particle distribution since there is  only
a factor $G_0$ which is different in the two calculations. This factor leads to a factor $-k^2/m$ in the present case, compared to the
particle distribution, that is we have for large $k$:
\begin{eqnarray}\label{eqintgdk1}
-\frac{1}{2i\pi } \int_{\mathcal C} d\omega \,f(\frac{\omega }{T})\,\left[\Sigma({\bf k},\omega)G({\bf k},\omega)
+\Delta ({\bf k},i\omega) F^{+}({\bf k},i\omega)\right]
\simeq -m\left[ R_{\Sigma}+|\Delta _{\infty}|^2\right]\,\frac{1}{k^2}
\end{eqnarray}
and the generalization of Eq.(\ref{eqen2}) is:
\begin{eqnarray}\label{}
E&=& 2  \sum_{\bf k} \left[\frac{k^2}{2m}\,n(k)-\frac{m \left[ R_{\Sigma}+|\Delta _{\infty}|^2\right]}{2\,k^2}\right] \\ \nonumber
&+& \sum_{\bf k}
\left[\frac{m \left[ R_{\Sigma}+|\Delta _{\infty}|^2\right]}{k^2}-\frac{1}{2i\pi } \int_{\mathcal C} d\omega 
\,f(\frac{\omega }{T})\,\left[\Sigma({\bf k},\omega)G({\bf k},\omega)
+\Delta ({\bf k},i\omega) F^{+}({\bf k},i\omega)\right]\right]
\end{eqnarray}

\section{CONCLUSION}
In this paper we have investigated the large momentum algebraic tail in the particle distribution and the energy
formula found by Tan associated with this tail. We have provided a simple derivation of this energy formula,
which rests on the fact that, in evaluating this energy for the short range potential under consideration, 
the interaction energy contribution is zero in most of phase space, which makes it negligible. 
Hence only the kinetic energy has to be calculated. The basis of the derivation is the careful subtraction of
the kinetic energy divergent contribution for interparticle distance less than the potential range.
This derivation is easily generalized to particles with different masses, to arbitrary mixtures, and
to two-dimensional space. We have then shown how the algebraic tail arises naturally in the field
theoretical many-body approach, from the analytic structure of the self-energy. Making use of these
ingredients we have shown how, starting from a standard general expression of the energy in terms
of the Green's function, one obtains a formula with a kinetic energy part which has the same structure
as in the formula found by Tan. This has been done both in the normal and in the superfluid state.
Finally we have taken the various exact formulas allowing to obtain the energy, and we have compared in the normal
state at unitarity the resulting numerical values obtained within the ladder approximation.

\section{ACKNOWLEDGEMENTS}
We are quite grateful for discussions to L. Tarruell and C. Salomon.
"Laboratoire de Physique Statistique de l'Ecole Normale Sup\'erieure" is "associ\'e au Centre National
de la Recherche Scientifique et aux Universit\'es Pierre et Marie Curie-Paris 6 et Paris Diderot-Paris 7".
We are also very grateful to D. d'Humi\`eres and N. Regnault for their recurring help in numerical matters.

\appendix

\section{Limiting cases}\label{A}

In the BEC limit $a \to 0_+$, one is led to consider the problem of a single molecule. The normalized wave function is
$\psi(r)=(2\pi a)^{-1/2}\,e^{-r/a}/r$ and the density distribution is given by the square of its Fourier
transform $n\us(k)=n\ds(k)=8\pi a^{-1}/(k^2+a^{-2})^2$. The coefficient of the $k^{-4}$ tail is $n_4=8\pi a^{-1}$.
One checks that indeed the energy $E=2 \sum_{k}(k^2/2m)(n(k)-n_4/k^4)+n_4/(4\pi ma)=-1/(ma^2)$ gives the
proper binding energy $E_b=1/(ma^2)$.

In the weak coupling limit $a \to 0_-$ the energy (per unit volume) can be expanded in powers of $k_Fa$
(the contributions which would come from BCS pairing are exponentially negligible). 
To second order \cite{agd}:
\begin{eqnarray}\label{enwc}
E=n \frac{k_F^2}{m} \left[\frac{3}{10}+\frac{k_Fa}{3\pi }+b_2 (k_Fa)^2\right]
\end{eqnarray}
where $n=k_F^3/3\pi ^2$ and $b_2=2(11-2 \log 2)/35\pi ^2 \simeq 0.0556613$. Second order perturbation
gives also:
\begin{eqnarray}\label{}
n\us(k)=n\ds(k)=\theta(k_F-k)+(k_Fa)^2\,{\tilde n}^{(2)}(k/k_F)
\end{eqnarray}
where $n^{(2)}(k)$ is a lengthy analytical expression given in Refs.\cite{bely,sar}. For large $k$, $(k_Fa)^2\,{\tilde n}^{(2)}(k/k_F) \simeq n^{(2)}_4/k^4$
with $n^{(2)}_4=(2k_F^3a/3\pi )^2$. Since there is no first order correction to the density distribution, the first order
correction to the energy, in the energy relation Eq.(\ref{eqfin}), is merely given by the last explicit term $n^{(2)}_4/4\pi ma=
\pi an^2/m$, which coincides indeed with the (mean field) first order term in Eq.(\ref{enwc}). 

To check this relation to second
order, we need the coefficient $n^{(3)}_4$ of the tail in the third order contribution $(k_Fa)^3\,{\tilde n}^{(3)}(k/k_F)$ to the density
distribution, which is not available. However we can extract it from the adiabatic relation \cite{brpl,tan2,zhlegg}
$n_4=4\pi ma^2 (dE/da)$, which gives $n^{(3)}_4=8\pi b_2nk_F^4a^3$. Hence we have to check that:
\begin{eqnarray}\label{}
2  \sum_{k} \frac{k^2}{2m}\left[(k_Fa)^2\,{\tilde n}^{(2)}(k/k_F)-\left(\frac{2k_F^3a}{3\pi}\right)^2\frac{1}{k^4}\right]
+\frac{2b_2nk_F^4a^2}{m}=b_2n \frac{k_F^2}{m}(k_Fa)^2
\end{eqnarray}
which implies:
\begin{eqnarray}\label{}
 \int_{0}^{\infty}dx\,\left[x^4\,{\tilde n}^{(2)}(x)-\frac{4}{9\pi^2}\right]=-\frac{2}{3}\,b_2
\end{eqnarray}
We have checked numerically this equation with a seven digits precision corresponding to all the digits given above for $b_2$.

Finally in the 2D case, in the BEC limit, one has again a single molecule problem. The normalized wave function of the bound
state with energy $E=-1/(2\mu \alpha ^2)$ is $\psi(r)=(\alpha \sqrt{\pi})^{-1}\,K_0(r/\alpha)$. For small $x$ the Bessel function
$K_0(x) \simeq \ln (2e^{-C}/x)$, so that $a=2\alpha e^{-C}$. The density distribution is  $n\us(k)=n\ds(k)=4\pi \alpha ^2/(1+(k\alpha )^2)^2$,
so that $n_4=4\pi/\alpha^2$. With these ingredients Eq.(\ref{eqfin2d}) is easily checked.

\section{Analytic structure of the self-energy}\label{B}

We discuss here in details the analytic structure of $\Sigma({\bf k},\omega)$ in order to justify the
assumptions we have made above in section \ref{normst}. It is useful to consider first the case of the ladder approximation \cite{agd,clk}, where the
structure is explicit. In this case the self-energy is given by:
\begin{eqnarray}
\Sigma ({\bf k}, i\omega_n ) = T  \sum_{{\bf K},\nu} \Gamma ({\bf K},i\omega _{\nu}) G_{0}({\bf K} - {\bf k},i\omega _{\nu}-i\omega_n)
=\frac{1}{2i\pi } \int_{\mathcal C} d\Omega \,b(\frac{\Omega }{T})\,
\sum_{{\bf K}}\Gamma ({\bf K},\Omega) G_{0}({\bf K} - {\bf k},\Omega -i\omega_n )
\label{sigmagen}
\end{eqnarray}
where in the last step we have replaced the summation over bosonic Matsubara frequencies $\omega _{\nu}=2\pi \nu T$,
with $\nu$ being an integer, by an integral over the contour $\mathcal C$ introduced above, with $b(x)=1/(e^{x}-1)$ 
being the Bose distribution function. The vertex $\Gamma ({\bf K},\Omega )$ is given by:
\begin{eqnarray}
\Gamma^{-1} ({\bf K},\Omega ) = \frac{m}{4\pi a} +  \sum_{{\bf k}'}\,
\left[ \,T  \sum_{m} G_{0\up}({\bf k}',i\omega _m) G_{0\down}({\bf K}-{\bf k}',\Omega -i\omega _m) -
 \frac{1}{2\epsilon _{k'}}\,\right]
\label{eqgam}
\end{eqnarray} 
Deforming contour $\mathcal C$ as we have done above in section \ref{normst}, we will from Eq.(\ref{sigmagen}) 
express $\Sigma ({\bf k}, \omega )$ in terms of the singularities of $\Gamma ({\bf K},\Omega )$
and of $G_{0}({\bf k},\omega )$, which are on the real frequency axis. 
On one hand $G_{0}({\bf k},\omega )$ has a simple pole at $\omega +\mu  =\epsilon _k $.
On the other hand the singularities of $\Gamma ({\bf K},\Omega )$ correspond first to the continuum of
scattering states with energy $\Omega +2\mu =\epsilon _{{\bf k}'}+\epsilon _{{\bf K}-{\bf k}'} \ge K^2/4m$,
arising from the product of the two $G_0$ in Eq.(\ref{eqgam}). In addition there is the possibility of a bound state
of the two particles,
corresponding to a zero of the right-hand side  of Eq.(\ref{eqgam}). As a result the spectrum of these singularities
has some lower bound $\Omega_{\rm min}$. On the other hand
for large $K$ their frequencies are bounded from below by $\Omega \approx  K^2/4m$, 
corresponding to the kinetic energy of the mass center.

Now let us first consider the contribution from the pole of $G_{0}({\bf K} - {\bf k},\Omega -i\omega_n )$ in Eq.(\ref{sigmagen}),
located at $\Omega=i\omega_n+\epsilon _{{\bf K} - {\bf k}}-\mu $. The Bose factor $b(\Omega /T)$ will produce a factor 
$f(\epsilon _{{\bf K} - {\bf k}}/T)$, which for large $k$ implies an exponentially small factor $e^{-k^2/2mT}$. 
This makes the corresponding contribution to the self-energy irrelevant for the algebraic tail of $n_k$. The only way to avoid this is to have
also a large ${\bf K} \approx {\bf k}$, so $\epsilon _{{\bf K} - {\bf k}}$ is not large. However in this case we will obtain
a factor $\Gamma ({\bf K},i\omega_n+\epsilon _{{\bf K} - {\bf k}}-\mu )$. The singularities of $\Sigma({\bf k},\omega)$
we are looking for are obtained from this factor by continuing the imaginary frequency $i\omega_n$ to the real $\omega $ axis.
However we know that, when $K$ is large, the singularities of $\Gamma ({\bf K},\Omega )$ are located at large frequencies
of order of $K^2/4m$. This implies that the corresponding singularities of $\Sigma({\bf k},\omega)$ will be for
$\omega+\epsilon _{{\bf K} - {\bf k}}-\mu \approx K^2/4m \approx k^2/4m$, that is $\omega \approx k^2/4m$
since $\epsilon _{{\bf K} - {\bf k}}$ is not large. As explained below Eq.(\ref{eqn}) these large $\omega $ singularities in
$\Sigma({\bf k},\omega)$ give only exponentially small contributions to $n_k$ and again do not come in the $k^{-4}$
tail we are looking for. Hence we conclude that the pole of $G_0$ in Eq.(\ref{sigmagen}) gives no contribution to
this tail, which comes accordingly only from the contribution of $\Gamma ({\bf K},\Omega)$.

For this contribution the dependence of $\Sigma ({\bf k}, \omega)$ on frequency comes explicitely from the $G_0$ term,
and it has poles for $\omega=\Omega -\epsilon _{{\bf K} - {\bf k}}+\mu $. Here $\Omega $ runs over the frequencies of the
singularities of $\Gamma ({\bf K},\Omega)$, but in practice the Bose factor limits them to some finite range $\Omega \lesssim T$.
This implies in particular that $K$ is bounded (just as above the terms produced by the tail of the Bose factor
will necessarily have an exponential factor and do not contribute to the algebraic tail of $n_k$). Hence, for large $k\rightarrow\infty$, we
obtain from the $G_0$ term a single pole located at $\omega \approx -k^2/2m$. The corresponding residue is, from Eq.(\ref{sigmagen}):
\begin{eqnarray}\label{eqresid}
R_{\Sigma}=-\frac{1}{2i\pi } \sum_{{\bf K}}\int_{\mathcal C_{\Gamma}} d\Omega \,b(\frac{\Omega }{T})\,
\Gamma ({\bf K},\Omega)=-\frac{1}{\pi }\sum_{{\bf K}}\int_{\Omega_{\rm min} }^{\infty} d\Omega \,b(\frac{\Omega }{T})\,
{\rm Im}\,\Gamma ({\bf K},\Omega+i\epsilon )
\end{eqnarray}
where $\mathcal C_{\Gamma}$ is a clockwise contour enclosing only the singularities of $\Gamma ({\bf K},\Omega)$
and the last expression (with $\epsilon \rightarrow0_+$) is the real axis integral obtained by calculating the contour integral from the jump of the imaginary
part of $\Gamma$ across the real frequency axis.

Let us consider now the general situation where no approximation is made. Actually, provided we replace the bare propagator
$G_0$ by the full propagator $G$, Eq.(\ref{sigmagen}) remains valid. We have:
\begin{eqnarray}
\Sigma ({\bf k}, i\omega_n ) = T  \sum_{{\bf K},\nu} \gamma ({\bf K},i\omega _{\nu};{\bf k}, i\omega_n) G({\bf K} - {\bf k},i\omega _{\nu}-i\omega_n)
=\frac{1}{2i\pi } \int_{\mathcal C} d\Omega \,b(\frac{\Omega }{T})\,
\sum_{{\bf K}}\gamma ({\bf K},\Omega;{\bf k}, i\omega_n) G({\bf K} - {\bf k},\Omega -i\omega_n )
\label{sigmagen1}
\end{eqnarray}
Indeed, considering the self-energy of, say, the $\up$ particle, there will be in any diagram a first interaction with a $\down$ particle,
hence correspondingly a $G \ds$ factor. Merely isolating this factor and gathering the rest into $\gamma$ produces Eq.(\ref{sigmagen1}).
We may express $\gamma$ in terms of the standard \cite{agd} full vertex ${\tilde \Gamma}$ by isolating the Hartree term in the self-energy.
This gives, for a contact interaction:
\begin{eqnarray}\label{}
\gamma(K;k)=g-g T\sum_{k'}G_{\up}(k')G_{\down}(K-k'){\tilde \Gamma}(k',K-k';k,K-k)
\end{eqnarray}
where we have used a four-vector notation $k \equiv ({\bf k},i\omega _n)$. Here the coupling constant $g$ is related to the
scattering length $a$ and to the cut-off $k_c$, related to the contact interaction, by
$g^{-1}=m_r/(2\pi a)- \sum_{}^{k_c}2m_r/ k^2$.

Physically $\gamma ({\bf K},i\omega _{\nu};{\bf k}, i\omega_n)$ describes the scattering of two particles $k$ and $K-k$,
and accordingly it will have the same qualitative properties as we have discussed above for $\Gamma ({\bf K},\Omega )$
(where we have been quite general for purpose). We consider the contributions in Eq.(\ref{sigmagen1}) from the singularities
of $G$. Either ${\bf K}$ is small compared to ${\bf k}$, so ${\bf K - k}$ is large and we can replace $G$ by $G_0$, and as above
we will have an exponential factor coming from the Bose factor. Or ${\bf K}$ is comparable to ${\bf k}$, in which case the corresponding
singularities of $\gamma$ will start around $K^2/4m$ and the resulting singularities for $\Sigma({\bf k},\omega)$ will be at least
$\omega \approx k^2/4m$, which makes them irrelevant for the algebraic tail of $n_k$.

Hence we have only to consider the contributions of the singularities of $\gamma$ in the calculation of Eq.(\ref{sigmagen1}).
Again those with large ${\bf K}$ (i.e. with small ${\bf K-k}$) will be located at high frequency $\Omega $, and the Bose factor 
produces an exponential factor which makes them irrelevant. We are left only with contributions arising for ${\bf K}$ small
compared to ${\bf k}$, in which case $G$ can again be replaced by $G_0$. This produces for $\Sigma({\bf k},\omega)$
an explicit pole at $\omega \approx -k^2/2m$, with a residue:
\begin{eqnarray}\label{}
R_{\Sigma}=-\frac{1}{2i\pi } \sum_{{\bf K}}\int_{\mathcal C_{\Gamma}} d\Omega \,b(\frac{\Omega }{T})\,
\gamma ({\bf K},\Omega;{\bf k}, -\frac{k^2}{2m})
\end{eqnarray}
Actually it seems physically reasonable that, in the large ${\bf k}$ limit, $\gamma ({\bf K},\Omega;{\bf k}, -k^2/2m)$
depends only on $({\bf K},\Omega)$. Indeed this limit corresponds to a short range  scattering between two $\up$
and $\down$ particles. In this case, in a way analogous to the one discussed in section \ref{detder} for the energy formula,
this scattering should be essentially described by two-body physics.
The situation is then similar to the one found in the ladder approximation.
The $k$ dependence is just the one which is already explicit, and we are left with a constant for the residue,
given by a formula like Eq.(\ref{eqresid}) with a modified $\Gamma ({\bf K},\Omega)$.

\end{document}